\documentstyle[sprocl,psfig]{article}

\def\Journal#1#2#3#4{{#1} {\bf #2}, #3 (#4)}

\def\NPB{{\em Nucl. Phys.} B}
\def\NPS{{\em Nucl. Phys. (Proc Suppl.)}}
\def\PLB{{\em Phys. Lett.}  B}
\def\PRL{\em Phys. Rev. Lett.}
\def\PRD{{\em Phys. Rev.} D}

\def\be{\begin{equation}}
\def\ee{\end{equation}}
\def\bea{\begin{eqnarray}}
\def\eea{\end{eqnarray}}
\newcommand{\eqn}[1]{Eqn.(\ref{eqn:#1})}
\newcommand{\fig}[1]{figure \ref{fig:#1}}
\newcommand{\tabl}[1]{table \ref{tab:#1}}
\newcommand{\order}[1]{{\cal{O}}(#1)}
\def\GeV{\;\mbox{GeV}}

\begin{document}

\title{The electroweak phase structure and baryon number violation for large Higgs mass.}

\author{ H.P.~Shanahan}

\address{DAMTP, University of Cambridge, 21 Silver Street,\\
Cambridge, CB3 9EW, England, U.K.}


\maketitle\abstracts{
The classical transitions between  topologically distinct 
vacua in a $SU(2)$--Higgs model,  using  
a Higgs field of mass approximately  $120 \GeV$, is examined 
to probe the crossover region between the symmetric and broken phase.
For the volumes used, this crossover is approximately $10 \GeV$ wide.}
  
\section{Introduction}

It is clear now from a number of numerical simulations in 3 and 4 
dimensions \cite{DESY-static-therm,Finn-static-therm}
that the first order phase transition from the broken to symmetric phase 
of the electroweak sector of the Standard Model
ends for a Higgs mass of approximately  $80  \GeV$. 
For larger Higgs masses, there is analytic crossover or very possibly 
a phase transition of third or lower order (whether numerical studies 
could ever distinguish between these two scenarios remains to be seen). 
While the present quoted lower bound  of $65  \GeV$ for the Higgs  
mass \cite{LEP-Higgs}
and an estimate for the Higgs mass from electroweak precision data
 \cite{ellis} is
$m_H =  145^{+164}_{-77}  \GeV$ 
is still consistent with an explanation of baryogenesis via this phase transition,
the amount of parameter space is rapidly shrinking. 

It seems logical to ask what is occuring to the real time process of
baryon number violation for larger Higgs masses.
Does the crossover become increasingly broad as the Higgs mass is increased
or does it remain reasonably sharp (changing from the symmetric to broken phase
over a few GeV) ?
An interesting scenario, discussed elsewhere in these 
proceedings \cite{Cosmology-Joyce}, suggests that
for a different cosmology at electroweak temperatures
one can still generate sufficient levels of baryon number violation, even with
crossover. At the very least, understanding baryon number violation
for this model will be important if we wish to employ a
similar mechanism  for more complicated field theories
such as SUSY or Grand Unified Theories.

\section{Real time transitions}
Many of the details of this calculation have been covered extensively in these
proceedings and so I will not go into detail here \cite{GM,AK,JS}.
Suffice it to say that we wish to examine the real time process of the
rate of change of the Chern-Simons number, which is related to baryon
number violation. While a full quantum field theory calculation
is almost impossible numerically, a classical treatment of the problem may give
a reasonable estimate for the rate.

\subsection{Classical analysis}
The following lattice $SU(2)$-Higgs Hamiltonian was constructed
\begin{eqnarray}
H \hskip-10pt &=& \hskip-10pt
 \beta\left[ - \sum_\Box\left(1 - \frac{1}{2}Tr U_\Box\right) -
\frac{1}{2} \sum_{x,i} ( \Delta_i \Phi )^\dagger ( \Delta_i \Phi ) \right. 
  \label{eqn:hamiltonian}  \\
 &-& \hskip-10pt\left.   \sum_x\left( \frac{M^2_{H0}}{2} \Phi^\dagger \Phi + 
\frac{\lambda_L}{4} (\Phi^\dagger \Phi)^2 \right) 
+ \frac{1}{(\Delta t)^2}\left( z_E \sum_{i,x} E_i E_i + \frac{z_\pi}{2} \sum_x 
\pi^\dagger \pi\right) \right]\nonumber
 \end{eqnarray}

The sums are defined over spatial sites only and the sum $\sum_\Box$ corresponds
to a sum over spatial plaquettes of the gauge field $U_i$. The first two terms in 
the Hamiltonian are discretized versions of $F^a_{ij}F^a_{ij}$ and 
$(D_i\Phi)^\dagger D_i \Phi$ respectively. The fields $E_i$ and $\pi$ are the conjugate
momenta to the coordinate fields $U_i$ and $\Phi$. 
The lattice spacing, $a$ has been taken to be 1.
Formally, \eqn{hamiltonian} results from the dimensional reduction of
the 4-dimensional $SU(2)$-Higgs Lagrangian with kinetic terms reintroduced. 
One would therefore  expect in the hamiltonian a Debye mass term and a renormalisation
of the kinetic terms. As noted by Moore and Turok \cite{MT1}, the Debye mass term
is effectively introduced by the U.V. cutoff of the lattice. The renormalisation
of the kinetic terms is expressed in the coefficient $z_E$ and $z_\pi$.
It is  believed that they take the form $1 + \order{g^2}$ and for this calculation
they are assumed to be 1. 
The time step factor $\Delta t$ is taken to be 0.05, which has been used in 
similar simulations.

A set of initial configurations are generated which satisfy 
Gauss constraints and then evolved classically using a leapfrog algorithm.
Instead of evaluating the Chern-Simons number directly, which have a number
of inherent problems, it is indirectly measured using a slave-field
method, proposed by Moore and Turok \cite{MT2}.  
This method has several advantages, being computationally quite
cheap and does not require a finite renormalisation.

One can set the Higgs mass by fixing the self-coupling term $\lambda_L$
since at tree level 
\begin{equation}
\frac{m_H^2}{m_W^2} \; \approx \; 2\lambda_L  \; ,
\end{equation}
for $m_H \approx 120 \GeV$,  $\lambda_L$ was set to 1.125.
In order to get an estimate of the temperatures, the relationships
between the 4-dimensional and 3-dimensional parameters 
can be used \cite{kajantie-4dto3d} 
to determine the ratio $m_H/T$ as a function of the 
remaining bare parameters, $\beta$ and $M_{H0}$.
For $m_H \approx 120 \GeV$ the temperature of crossover, $T_{cross}$ is around
$200 \GeV$ \cite{Finn-static-therm}.
Hence, $M_{H0}$ was set to -0.596 so that $aT_{cross}$ corresponded to 
$\beta \approx 7.5$ (which is inversely proportional to $aT$). 

\begin{center}
\begin{figure}
\hbox{
\psfig{figure=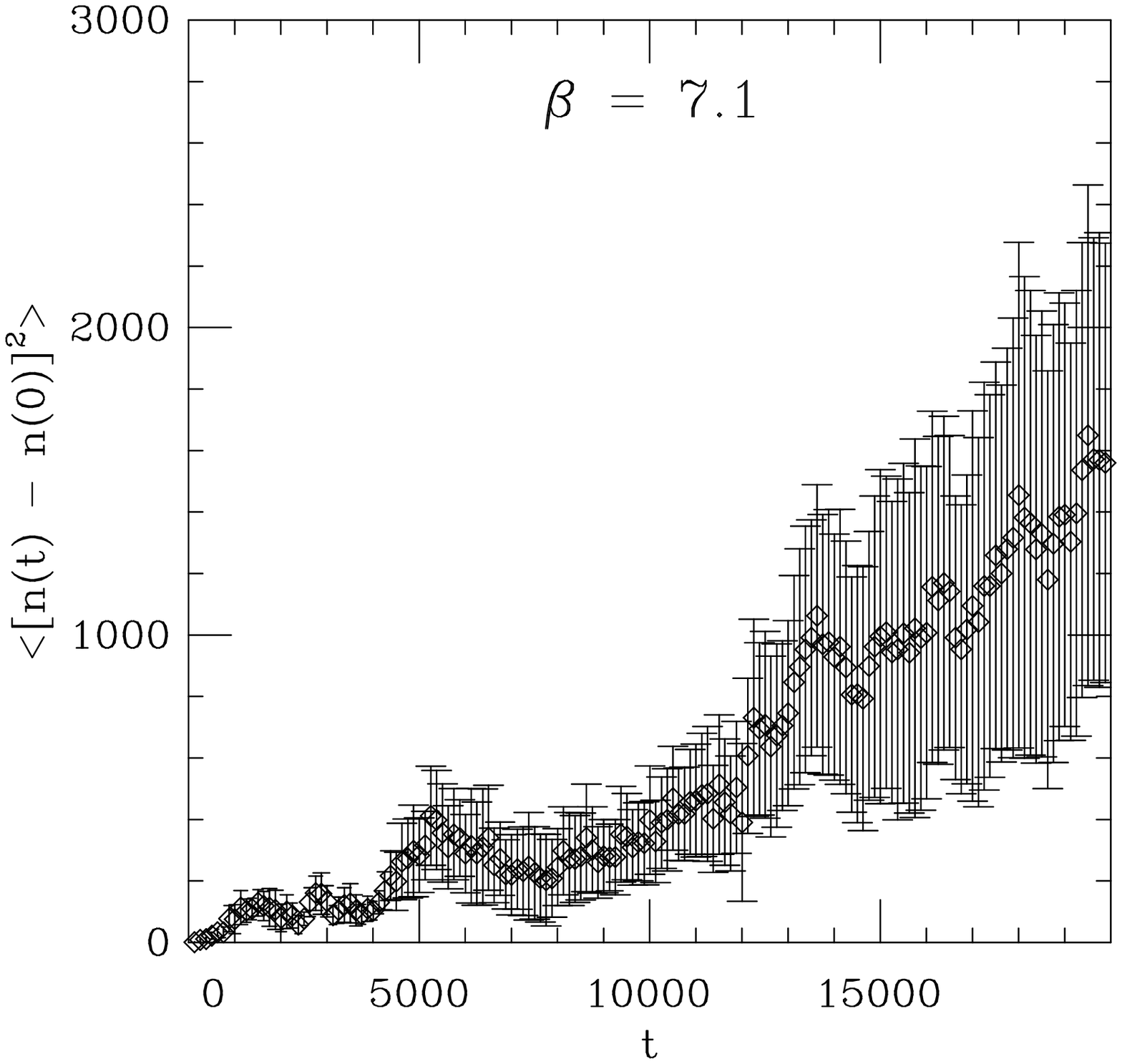,height=2.2in}
\psfig{figure=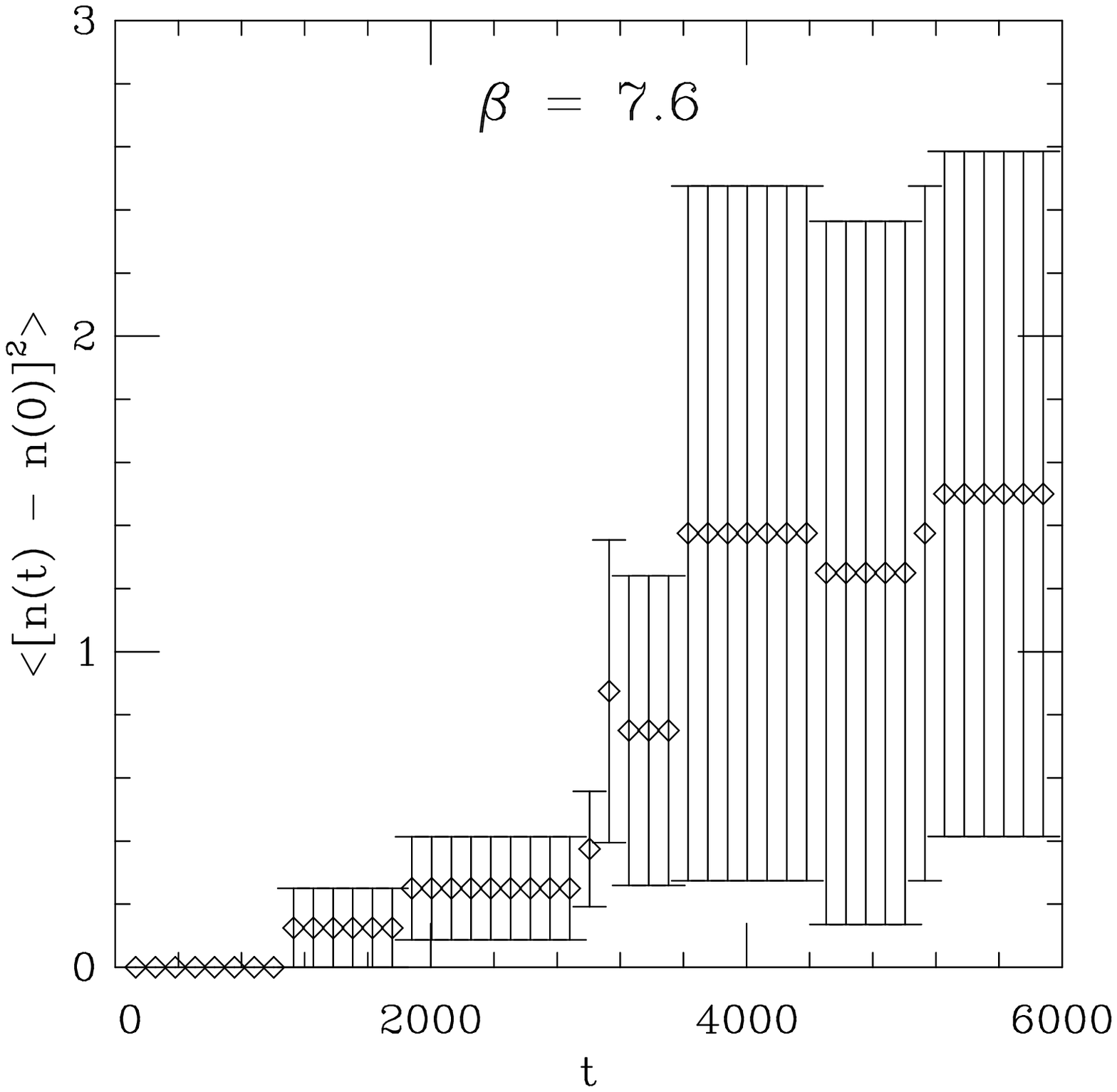,height=2.2in}}
\caption{Diffusion rate of winding number at $\beta=7.1$ and $\beta=7.6$. Note
the difference in the vertical scales.
\label{fig:random_walk}}
\end{figure}
\end{center}

\subsection{Results}
\begin{table}[t]
\vspace{0.4cm}
\begin{center}
\begin{tabular}{|c|c|c|c|}
\hline
$\beta$ & \#configurations & \#time steps & $\kappa$ \\ \hline
6.8 & 16 & 7000 & 0.973 $\pm$ 0.057 $\pm$ 0.161 \\
6.9 & 16 & 7000 & 0.984 $\pm$ 0.061 $\pm$ 0.107 \\
7.0 & 16 & 10000 & 1.013 $\pm$ 0.083 $\pm$ 0.187 \\
7.1 & 8 & 20000 & 0.770 $\pm$ 0.059 \\
7.2 & 8 & 10000 & 0.388 $\pm$ 0.034 $\pm$ 0.038\\
7.3 & 8 & 10000 & 0.092 $\pm$ 0.008 \\
7.4 & 8 & 6000 & 0.018 $\pm$ 0.002 $\pm$ 0.003\\
7.5 &  8 & 9000 & 0.00787 $\pm$ 0.00068 \\
7.6 & 16 & 6000 & 0.00088 $\pm$ 0.00006 \\
7.7 & 8 & 6000 & 0.00130 $\pm$  0.00016\\
7.8 & 8 & 9000 & 0.00055 $\pm$  0.00005\\  
 \hline
\end{tabular}
\end{center}
\caption{Numbers of configurations and the time steps iterated
forward for each $\beta$ \label{tab:sweeps}. The second error
quoted for $\kappa$ is the difference between the first and second bins
of 50 coefficients when the difference was greater than 1 standard deviation.}
\end{table}

\begin{center}
\begin{figure}
\hbox{
\psfig{figure=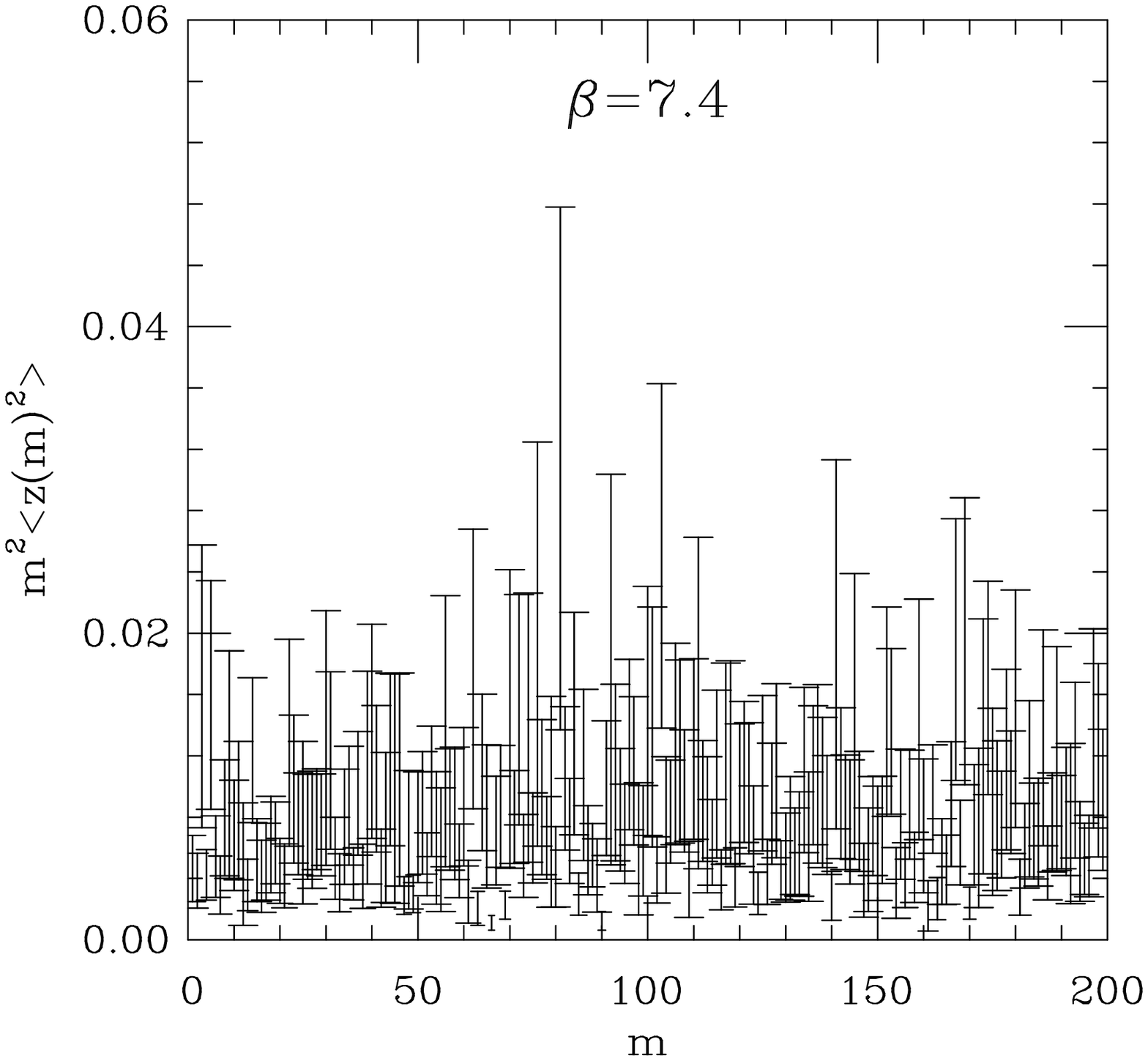,height=2.2in}
\psfig{figure=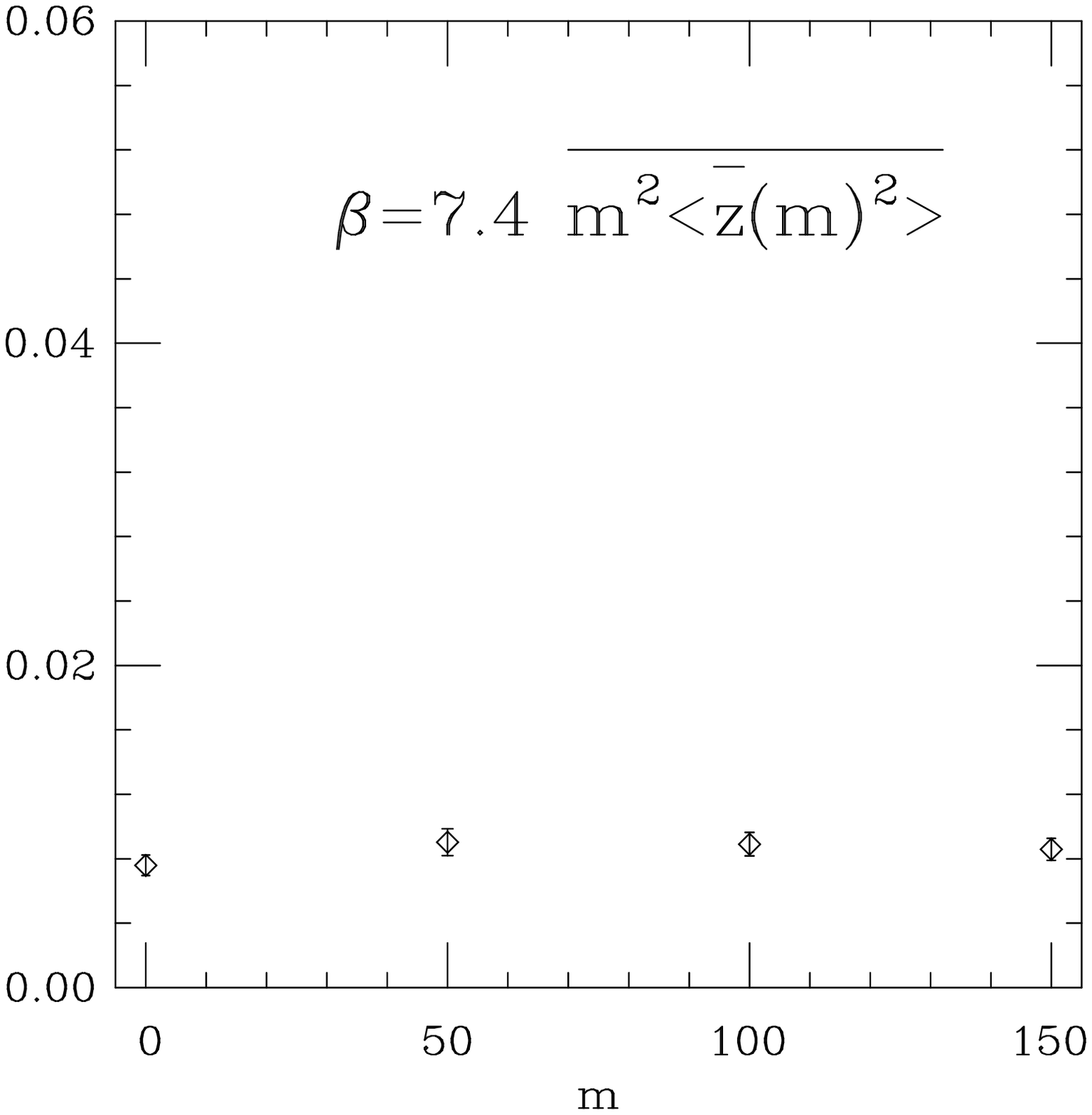,height=2.2in}}
\caption{The mean square of the cosine coefficients at $\beta=7.4$ and the average 
of the coefficients with a bin size of 50.
\label{fig:cosine}}
\end{figure}
\end{center}

For a lattice size of $24^3$, $\beta$ was varied from 6.8 to 7.8.
The number of initial configurations and number of sweeps are shown 
in \tabl{sweeps}. As can be seen from \fig{random_walk}, one could determine
the diffusion coefficient directly from the data by evaluating the slope.
However, in order to maximise the sample size, a cosine 
transform was applied \cite{MT2} to the
data, and the first 200 coefficients evaluated.
The diffusive contribution to the coefficient will be proportional to $1/m^2$, 
where $m$ is the number
of the coefficient. 
An example of the
resulting coefficients is shown in \fig{cosine}. 
The error for each coefficient
is evaluated via jack-knife. 
Correlations between different coefficients appear to drop significantly
as the statistics is improved with the percentage of correlation coefficients
greater than 0.9 dropping to less than 5\% for the larger data sets.
The coefficients
were averaged in bins of 50, which reduces the error significantly. 
The errors were summed quadratically. (One could also evaluate the
error for the binned data by treating the central values of the 
coefficients  as independent points and estimating the error as a standard deviation.
This approach varied the statistical error by a factor of 0.5 to 2. Ideally, one should perform
a correlated fit through these points, however the sample size is not large enough
to do this.)
For some values of $\beta$, differences between different regions of binning
varied by more than one standard deviation. A conservative 
estimate of the systematic error due to fitting took this 
into account.

\section{Conclusions}
\begin{center}
\begin{figure}
\hbox{
\hskip3truecm
\psfig{figure=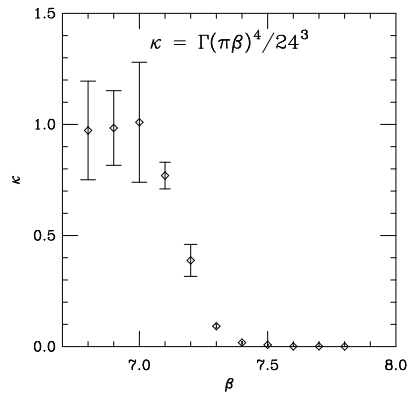,height=2.2in}
}
\caption{$\kappa$ as a function of $\beta$ using a $24^3$ lattice.
\label{fig:kappa_vs_beta}}
\end{figure}
\end{center}

From \fig{kappa_vs_beta} one can see that the  diffusion rate vanishes
for $\beta > 7.4$.
The diffusion rate is decreasing from $\beta=7.0$. 
This corresponds to a temperature variation of around 5\%,
which, assuming the crossover temperature is around $200 \GeV$ implies a variation
of about $10 \GeV$. The rate for $\beta=7.0$ is of the same order as similar simulations
done for $m_H \approx m_W$ \cite{Tang-Smit,AK,MT2}.
The calculation of $\kappa$ for $\beta \; < \;7.0$ is consistent with 
$\kappa(\beta=7.0)$; however
the Moore-Turok definition of the 
winding number will very often change by more than 1 between measurements which
introduces a systematic error.  The next step is to adjust $M_{H0}$ so that 
$aT_{cross}$ is much smaller and thereby larger $\beta$'s (smaller lattice spacings) 
are  required. 
A study of the effect of the finite volume is also necessary.
Ultimately this should be repeated for a larger Higgs mass ($150 \GeV$, for example) to
determine if the width is increasing with the Higgs mass. 

\section*{Acknowledgments}
I am indebted to Guy Moore and Neil Turok for making 
their code available for this simulation.
I thank Anne Davis, Ron Horgan, Alex Krasnitz and Jan Smit for useful
conversations.
This research is supported by the Leverhulme Foundation.
The numerical work  was carried out on a Silicon Graphics/Cray
Origin 2000 run by the UK Computational Cosmology Consortium
and an  Hitachi SR2201 at the
High Performance Computing Facility  at the University of Cambridge. 
Support for these computational facilities has been provided by Silicon Graphics/Cray,
Hitachi, HEFCE, PPARC and the University of Cambridge.

\end{document}